\documentclass[reprint,amsmath,amssymb,aps,superscriptaddress ]{revtex4-1}

\usepackage{amsmath,amssymb,amsfonts}
\usepackage{algorithmic}
\usepackage{graphicx}
\usepackage{epsf}
\usepackage{epsfig}
\usepackage{textcomp}
\usepackage{xcolor}
\usepackage{float}
\usepackage{ulem}  

\begin{document}

\title{Charge and heat currents in prismatic tubular nanowires}

\author{Hadi Rezaie Heris}
\affiliation{Department of Engineering, Reykjavik University, Menntavegur 1, IS-102 Reykjavik, Iceland}

\author{Kristjan Ottar Klausen}
\affiliation{Department of Engineering, Reykjavik University, Menntavegur 1, IS-102 Reykjavik, Iceland}

\author{Anna Sitek}
\affiliation{Department of Theoretical Physics, Wroclaw University of Science and Technology, Wybrze\.{z}e  Wyspia\'{n}skiego  27, 50-370 Wroclaw, Poland}

\author{Sigurdur Ingi Erlingsson}
\affiliation{Department of Engineering, Reykjavik University, Menntavegur 1, IS-102 Reykjavik, Iceland}

\author{Andrei Manolescu}
\affiliation{Department of Engineering, Reykjavik University, Menntavegur 1, IS-102 Reykjavik, Iceland}

\begin{abstract}
We calculate electronic charge and heat transport in tubular nanowires generated by a temperature gradient or a chemical potential bias. These nanowires correspond to semiconductor core-shell nanowires with insulating (undoped) core and conductive (doped) shell, such that the conduction takes place only in the shell.  The cross section of such nanowires is typically polygonal.  We study the influence of the cross section shape and shell thickness on the electric and heat conduction of the shell. We use the Landauer-B\"uttiker approach to calculate the electric and heat currents as a non-linear function of temperature and chemical potential bias beyond the linear regime.
\end{abstract}

\maketitle

\section{Introduction}

Nanowires made of different materials allow several controllable properties of the electronic states within channels of widths from tens to hundreds of nanometers.  Such structures can be built as radial heterojunctions of two different semiconductor materials, and are known as core-shell nanowires \cite{blomers2013realization,shi2015emergence}. Such nanostructures have a variety of electronic properties, determined by the band alignment between the two materials, by the size of the core, or by the shell thickness or shape, which  
make them attractive for a variety of applications of quantum devices such as nanosensors, solar cells, optical detectors or thermoelectric convertors \cite{pistol2008band,tang2011solution,peng2015single,Dominguez19}. 

Core-shell nanowire based on III-V semiconductors are almost always prismatic, with a polygonal profile. The typical cross section shape is hexagonal \cite{rieger2012molecular}, although other prismatic geometries such as square and triangular have also been fabricated \cite{qian2005core,Goransson19}. Interestingly, it is also possible to remove the core, and to  obtain hollow nanowires, with vacuum inside \cite{rieger2012molecular}.  Because of these many possibilities, understanding the implications of the geometry on electronic properties such as electric and heat conductance is important. 

In the present work we consider models of shells with several geometries: cylindrical, hexagonal, square, and triangular, with lateral thickness of up to 40\% of the overall diameter of the nanowire, and we calculate the electric and heat current of electrons in the presence of a temperature gradient and chemical potential bias. A temperature gradient across a conducting material induces an energy gradient which leads to charge or energy transport. The particles on the hotter side have larger kinetic energy than those at the colder side, and therefore the net particle flow is expected to be from the hotter to the colder side. 
However, the thermoelectric current can be positive or negative, depending on the type of carriers, i.e.\ electrons or holes, or more generally, depending on the occupation fraction of the transverse modes contributing to the transport \cite{erlingsson2017reversal}. 

For a tubular nanowire with circular symmetry the electrons have uniform angular distribution within the shell. In the case of polygonal shells the localization pattern, especially for low-energy states, is much more complicated \cite{Sitek15,sitek2016multi}. Consequently, the properties of polygonal nanowires may differ considerably from their circular counterparts.  In general, the lateral confinement of electrons in the tubular shell leads to an increase in thermoelectric current compared to the values in the bulk materials \cite{heris2020thermoelectric}. At the same time the thermal conductivity due to phonons can be strongly suppressed in nanowires with a diameter below the phonon mean free path \cite{boukai2008silicon,Dominguez19}. Therefore semiconductor core-shell nanowires provide a unique opportunity to design desired nanoscale devices by engineering the charge and heat transport related only to electrons, through the cross section area configuration. Depending on the purpose, suppressing heat and increasing charge current, or vice versa, can be achieved via geometric elements \cite{schelling2005managing,zhao2014ultralow}. 
By considering electronic conductance features and variations as function of transverse geometry and shell thickness, and the transported heat through the core, one can calculate the figure of merit or the thermoelectric efficiency, and find optimal situations. Note that for this purpose the heat transported by phonons and electron-phonon interaction must also be considered \cite{heris2022effects,tsaousidou2008energy}.

\section{Model and methods}

\begin{figure}
  \centering
  \includegraphics[trim = 0 16 215 45, clip, width=0.48\linewidth]{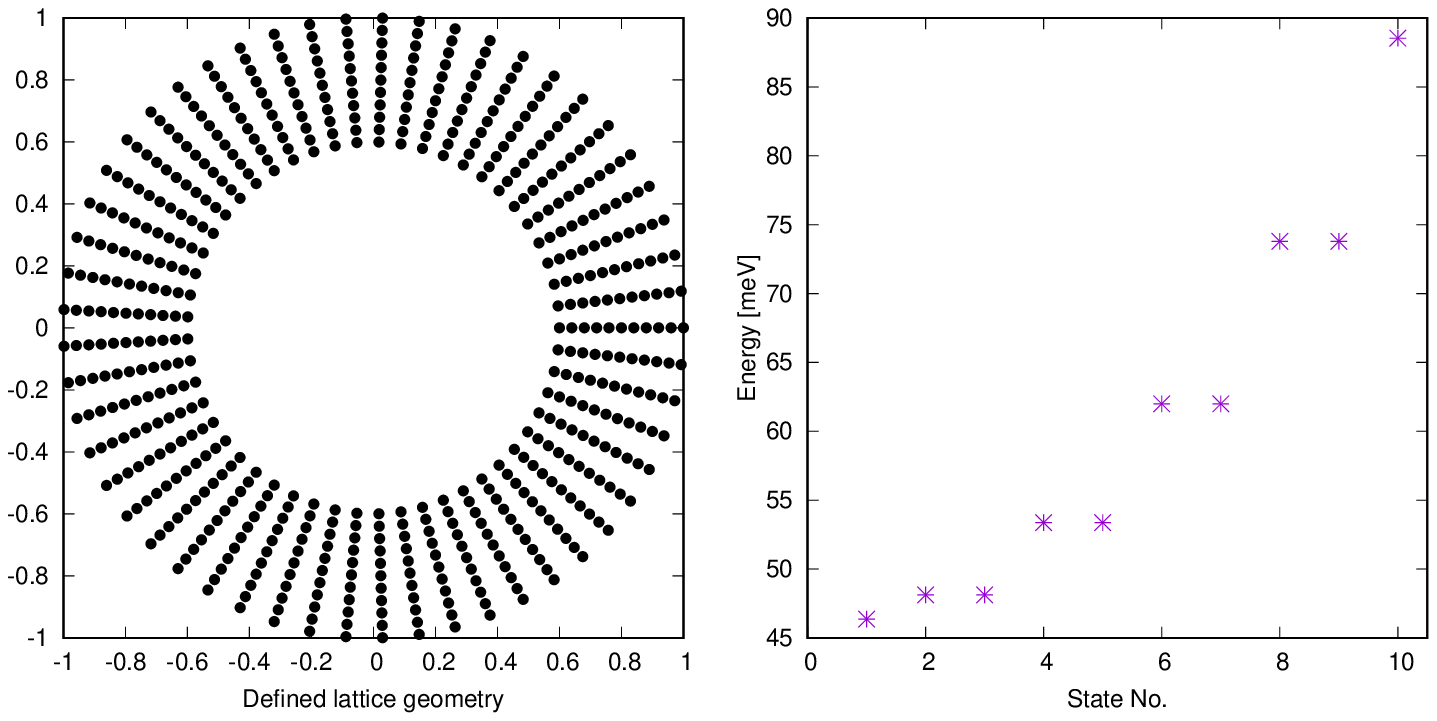} 
  \includegraphics[trim = 0 16 215 45, clip, width=0.48\linewidth]{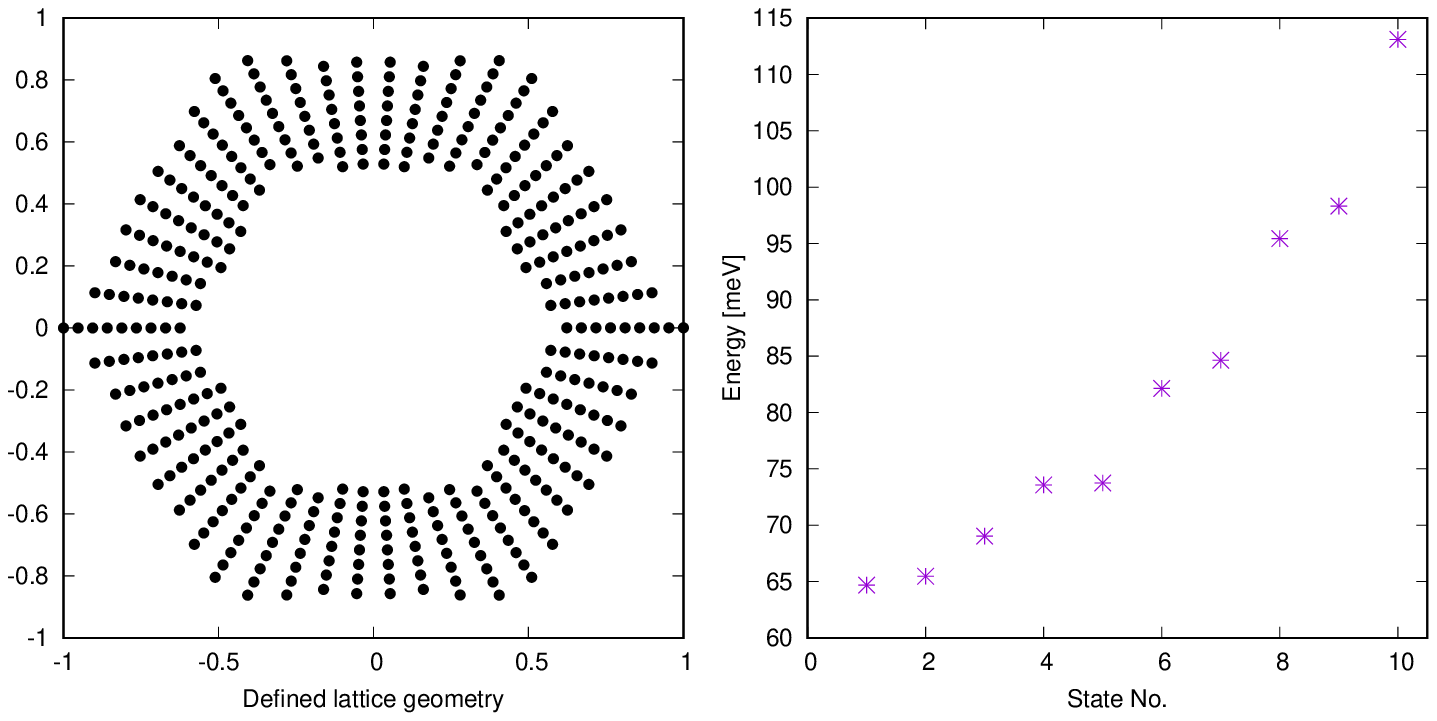}
  \includegraphics[trim = 0 16 215 30, clip, width=0.48\linewidth]{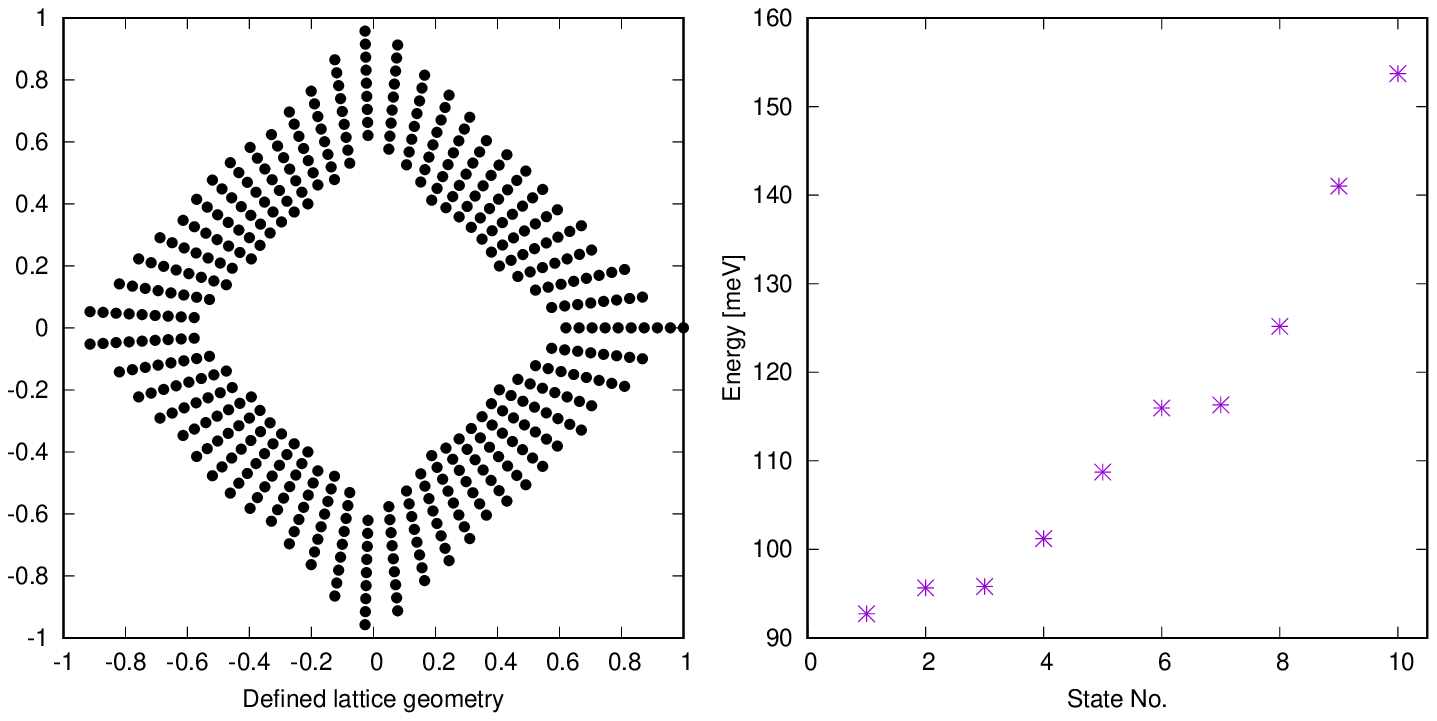}
   \includegraphics [trim = 23 13 53 0, clip, width=0.48\linewidth]{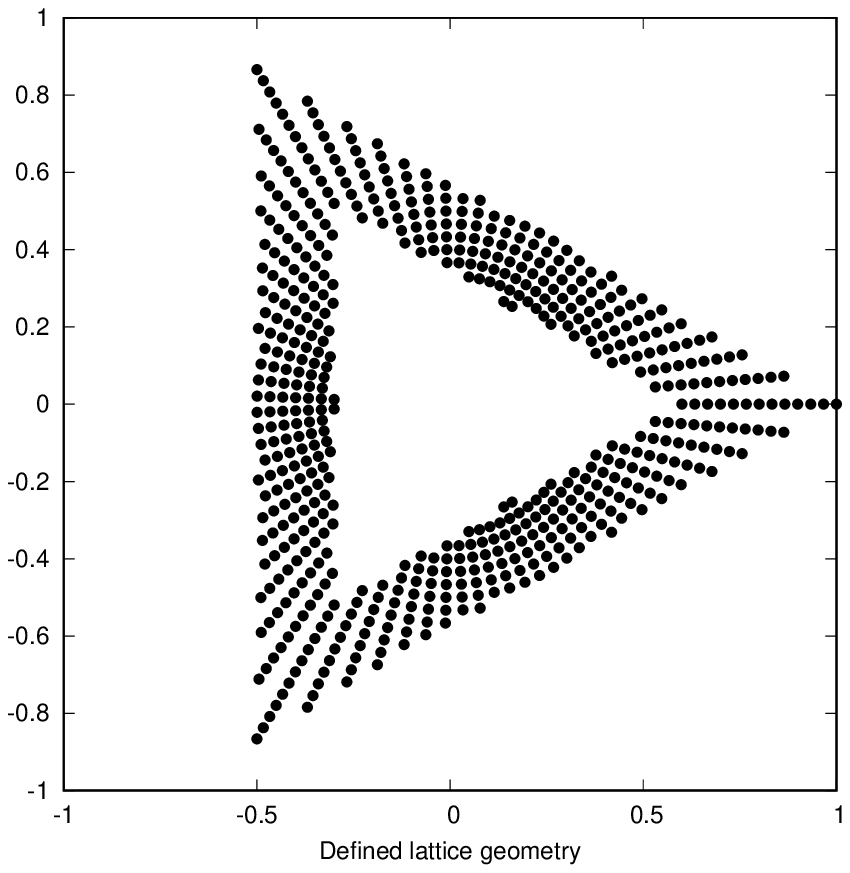}
  \caption{The discretized polygonal with external radius of 50\,nm and thickness of 20\,nm. the cross section of the prismatic shells are defined by applying boundaries on a circular ring discretized in polar coordinates-points indicate the shell thickness.
  }
\label{fig7}
\end{figure}

Our model is a systems of electrons confined in prismatic shell with cylindrical, hexagonal, square
or triangular cross section. In all cases we consider an external radius of the polygonal shell of 50\,nm (i.~e. the radius of the circle encompassing the entire cross section), and initially the side thickness will be 20\,nm.  The Hamiltonian can be decomposed into a transverse term and a longitudinal term. The transverse Hamiltonian is defined on a lattice of points that cover the cross section of shell. For this purpose first we define a radial lattice within a circular disk on which we integrate the desired polygonal shell and exclude from the Hamiltonian the lattice points which are outside of the shell \cite{daday2011electronic,torres2018conductance}. The transverse geometry of our nanowire models is presented in Figure~\ref{fig7}.
The discretized ring was represented by 10-15 radial $\times$ 40-54 angular sites, depending on the polygon shape and shell thickness.  The electron effective mass is as for bulk InAs, $m_{\mathrm {eff}}=0.023\,m_{\mathrm e}$.  

%

We calculate the charge current, $I_{c}$, and heat current, $I_{q}$,  through the nanowire using Landauer-B\"uttiker approach, based on the transmission function $\mathcal{T}(E)$:
%
 \begin{equation}
     I_{c}=\frac{e}{h} \int 
     \mathcal{T}(E)[f_{L}(E)-f_{R}(E)] \, dE \ ,
 \end{equation}
\begin{equation}
     I_{q}=\frac{1}{h} \int 
     \mathcal{T}(E)[E-\mu][f_{L}(E)-f_{R}(E)] \, dE \ ,
\end{equation}
where 
\begin{equation}
f_{L,R}(E)=\frac{1}{1-e^{(E-\mu_{L,R})/k_{B}T_{L,R}}} \ 
\end{equation}
is the Fermi function for the left ({\it L}) or right ({\it R}) reservoirs with the chemical potential $\mu_{L,R}$ and temperature $T_{L,R}$. 

We consider a fixed chemical potential bias $\Delta \mu=\mu_{L}-\mu_{R}$ and change the values of the chemical potentials simultaneously, at each end of nanowire, in each step. For studying the effect of the cross section shape, the values of the chemical potential at the left reservoir start from 45\,meV, which is the minimum value corresponding to the circular cross section, and goes up to 225\,meV, which is the highest energy level considered in the triangular cross section. For studying thickness effects, the left reservoir chemical potential range will be 140-850\,meV, for nanowires with thicknesses of 20\,nm and 5\,nm.  Because of the different geometries, the energy spectra are different. In each case we shall consider the first ten transverse modes to be inspected with the chemical potential window.

We consider a fixed temperature gradient $\Delta T=T_{L}-T_{R}=35$\,K (which is very close $k_{B}T$ value) and we change both ends temperatures at the same time in each step. In the first step, $T_{L}=36$\,K and $T_{R}=1$\,K and simultaneously we increase each reservoir temperature by 35\,K in each step. There are 12 steps for each of the charge and heat currents in both cases of cross section and thickness calculations.
The energy integral is calculated numerically using the trapezoidal method. 

We calculate the charge and heat currents $I_{c}$ and $I_{q}$ by varying one parameter at a time.
Note that we neglect carriers interaction and scattering with impurities, and consider only ballistic transport of electrons in the nanowires. In this case the transmission function $\mathcal{T}$ is a multistep function of energy. The presence of disorder indeed changes the transmission function, but in the present work we neglect disorder, or assume it is sufficiently weak such that disorder effects are dominated by the thermal effects. 

\section{Results and discussion}


\subsection{Effect of transverse geometry}

In shells with polygonal cross section area, the lowest energy states are localized along the corners, and there is a considerable energy gap between the states localized at corners and the next group of states localized on polygon sides.   This gap increases with decreasing the shell thickness and with decreasing the number of corners \cite{Sitek15}. However, experimentally, this localization effect has not been thoroughly investigated.  Whereas for the electrical conductance without a temperature bias, i.~e. when $T_L=T_R$, the conductance is expected to have steps corresponding to the transverse modes \cite{torres2018conductance}, the thermoelectric and heat transport are different.
\begin{figure}
\centering
\includegraphics[width=1\linewidth,height=46mm]{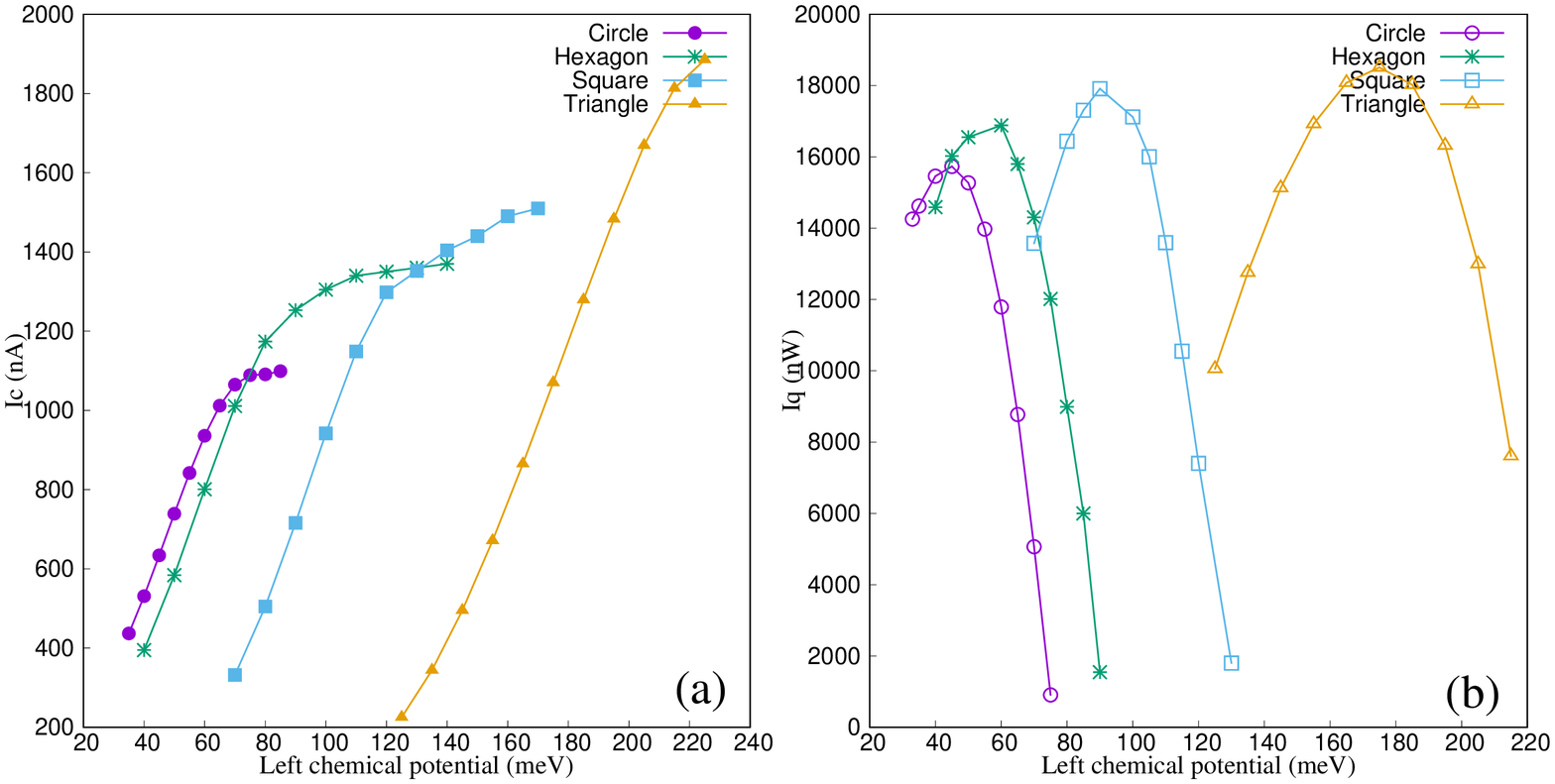}
\includegraphics[width=1\linewidth,height=46mm]{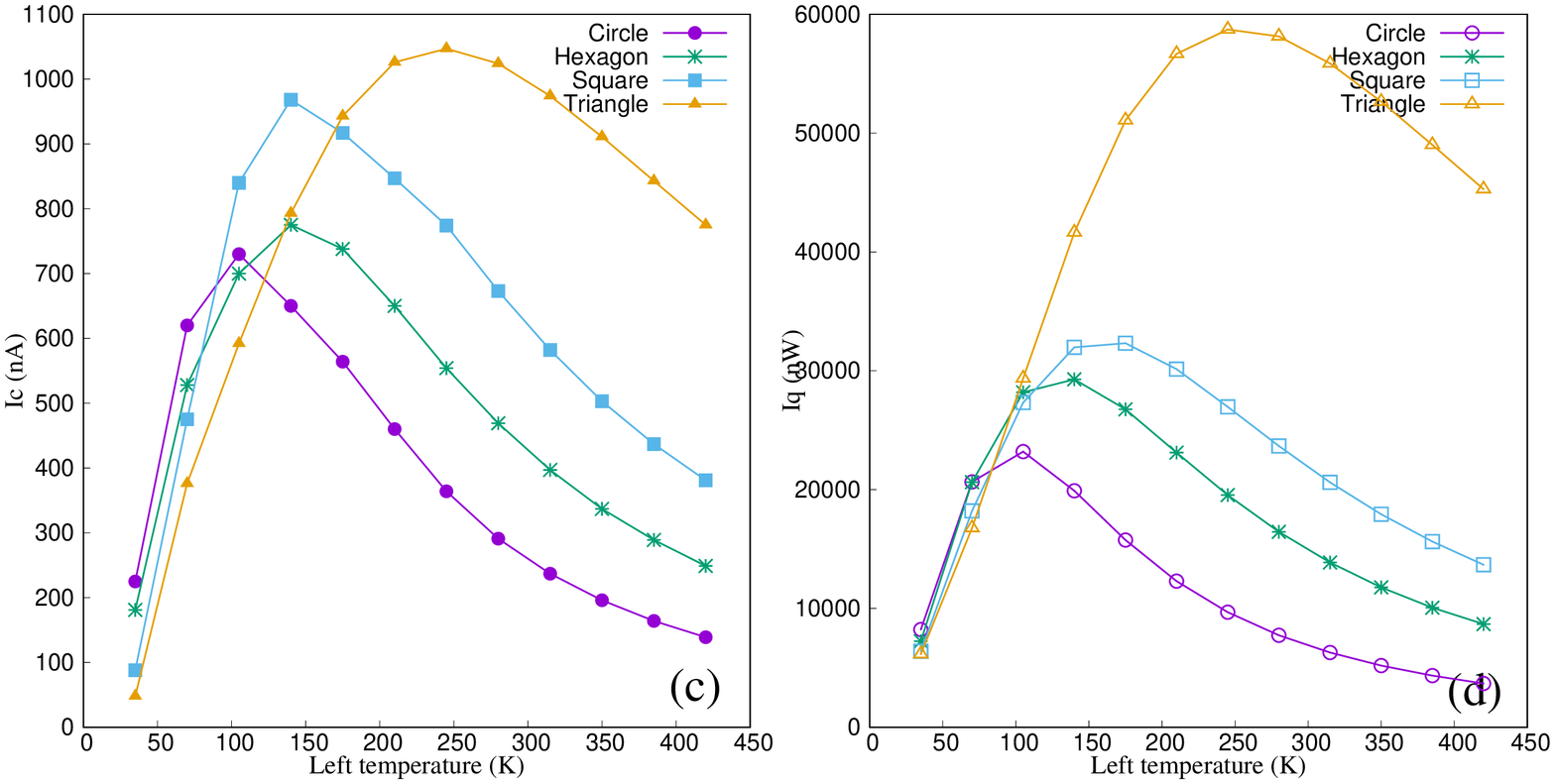}
 \caption{Cross section shape effect on electrical current and heat current as function of left reservoir chemical potential (a) and (b) and temperature (c) and (d). (a) and (c) are representative of electrical current and (b) and (d) are for heat current. In case of chemical potential bias, $T_L=T_R=200$\,K and in presence of temperature gradient chemical potentials values calculated as mentioned in text.}
\label{currents} 
\end{figure}
%
Figure~\ref{currents} presents electrical and heat currents as function of left reservoir chemical potential and temperature for different cross sectional shapes of our tubular nanowires. In Figure~\ref{currents}(a) we can see that the electrical current increases with increasing the chemical potential for all shapes, and that is reaches larger values for the polygonal geometry than for the circular case. Although for each cross section the energy states are different, the charge current in all cases simply follow the transmission function and the energy window associated with the chemical potentials. By increasing the chemical potentials more states will participate in transmission which lead to a higher charge current flow.  The largest electric current corresponds to the triangular cross section. These values for electrical conductance are obtained for all shapes without a temperature bias, $T_L=T_R=200$\,K,  and as a function of chemical potential bias.
Another interesting issue about electrical current variation with the chemical potential for different cross sectional shapes is that in all cases electrical current will be saturated at high values of chemical potentials. 

The heat current variation with increasing the chemical potential shows a peak for all shapes but we can also see a rapid decrease after that peak, Figure~\ref{currents}(b). The position of the peaks is related to the fixed temperatures of both ends, i.e.\ if we choose both temperatures 300\,K instead of 200\,K these peak shift a little to the right side. The reason for this shift is that when we increase the temperature, $k_{B}T$ will increase and the system needs a higher chemical potential to activate the higher number of carriers.     

For calculating the temperature effect on the electrical conduction we change left and right reservoir temperatures as mentioned in Section II. Now there is no chemical potential bias in the system, and for each geometry the chemical potentials of the left and right reservoirs are considered equal, $\mu_{L}=\mu_{R}=(\mu_{max}-\mu_{min})/2$. For the circlular cross section $\mu_{max}=88$\,meV and $\mu_{min}=46$\,meV, so we put both chemical potentials at 67\,meV. The same had been applied to the other shapes and chemical potential values are 80\,meV, 130\,meV and 183\,meV, for the hexagon, square, and triangle, respectively. 
In Figure~\ref{currents}(c), by increasing the temperature of the left lead, 
the electric current initially increases, but after reaching a maximum it decays slowly. 
The maximum values of the charge current for each shape occur at different temperature. The lowest peak occurs in the circular case, and the highest peak for the triangular geometry. The highest value of the charge current for triangle shape occurs at 250\,K, which is a quite high temperature regime with respect to the other geometries. Also, the charge current values at high temperature are nearly two or three time larger than for other shapes. 
At low temperatures the charge current for polygonal shapes with fewer corners is lower, while by increasing the temperature we see the reverse situation, i.~e. polygonal shapes with fewer corners show higher charge current. 
The heat current variation with the left temperature shows the same behavior up to 100\,K 
for all shapes, but above that temperature again the triangular case has the highest current and the circular the lowest values for the transported heat. Just like the electric current, heat current has a peak occurring different temperatures for different cross sectional shapes, and after that it decreases smoothly to a saturated value.





%
 \begin{figure}[t]
     \centering
     \includegraphics[trim = 50 1 45 5, clip, width=0.48\linewidth]{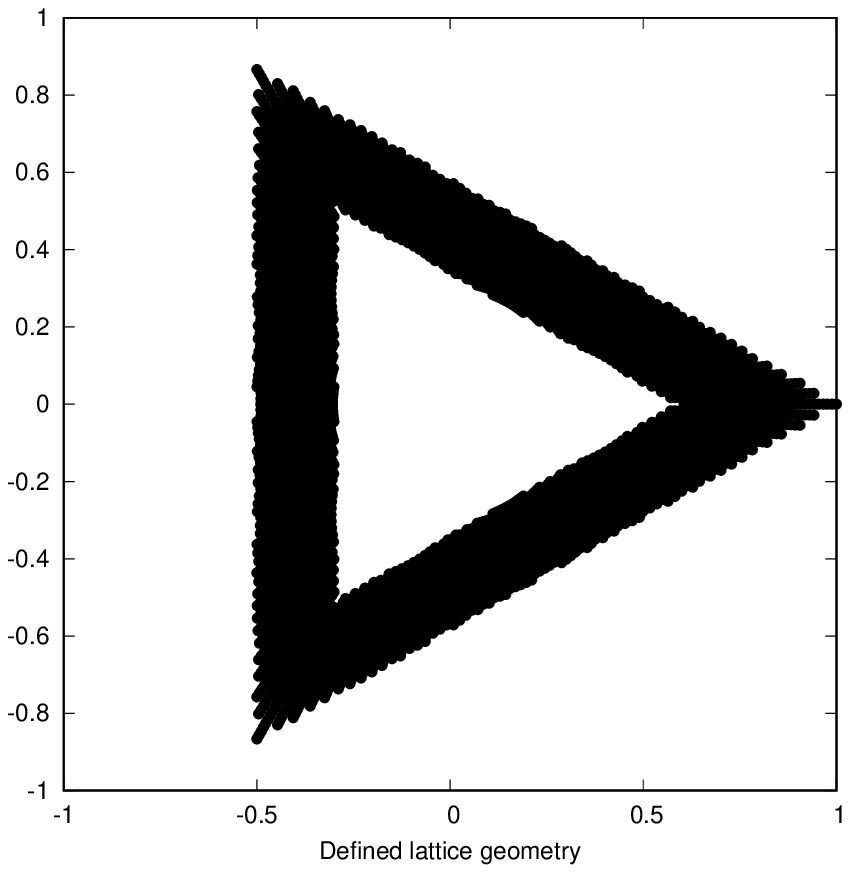}
     \includegraphics[trim = 230 5 0 33, clip, width=0.48\linewidth]{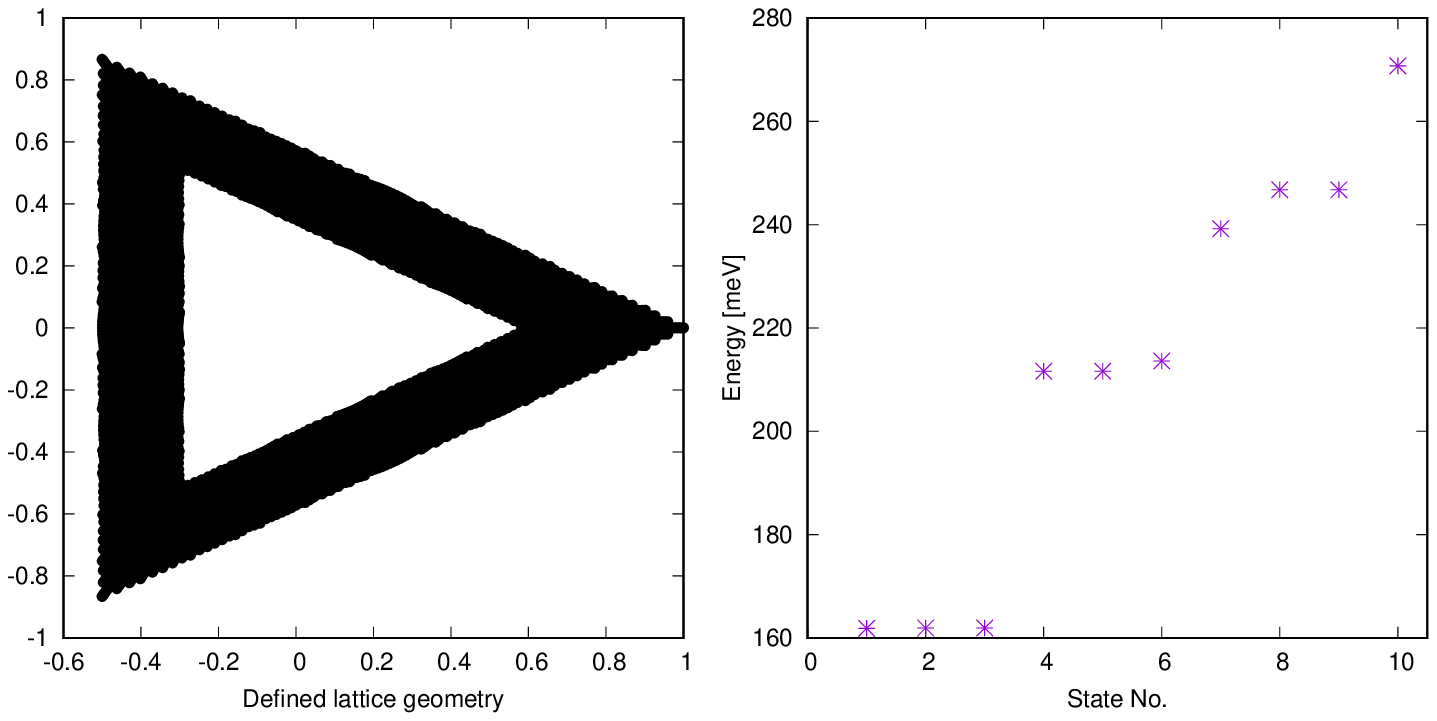}
     \caption{The energies of the transverse modes for a triangular shell of 20~nm thickness and 50~nm external radius, are shown in the right figure.  There are three states at $\approx 160$~meV (spin degenerated), localized in the corners, and six states at $\approx 212-215$~meV localized on the sides.  The energy gap between these states increases when the shell thickness decreases.}
     \label{energy_spectrum}
 \end{figure}

\subsection{Effect of thickness}

Next we shall consider the triangular cross sectional shape for studying the thickness effect on the electronic charge and heat transport. The working samples will have a fixed external radius and variable side thickness.
The energy levels strongly depend on the aspect of ratio of the polygonal cross section, i.~e. the ratio of thickness to diameter, in a way that the energy differences increase when this ratio is small. In particular, for  the triangular case the energy gap between states localized at corners and states localized on sides strongly increases with decreasing the side thickness and keeping the total thickness fixed \cite{sitek2016multi}.  An example is shown in Figure~\ref{energy_spectrum}.  In the next calculations we shall use four shell thicknesses,  5, 10, 15, and 20\,nm and constant external radii 50\,nm. Similar to the calculations for variable geometry, in this part we consider $T_{L}=T_{R}=200$\,K in case of a chemical potential bias. And for a temperature bias we consider left and right chemical potentials as average of minimum and maximum of chemical potentials for that specific thickness.

\begin{figure}
\centering
\includegraphics[width=1\linewidth,height=46mm]{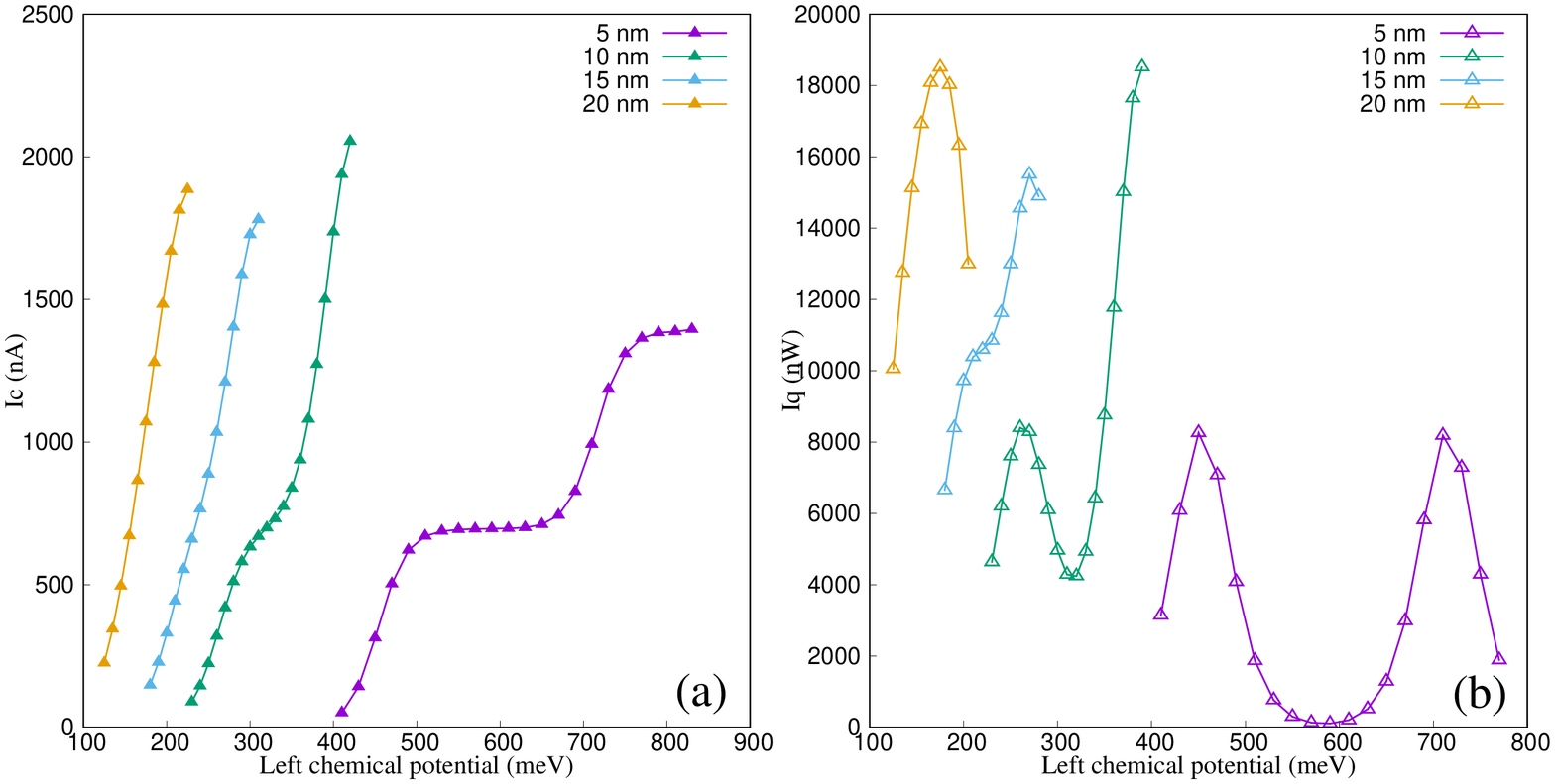}
\includegraphics[width=1\linewidth,height=46mm]{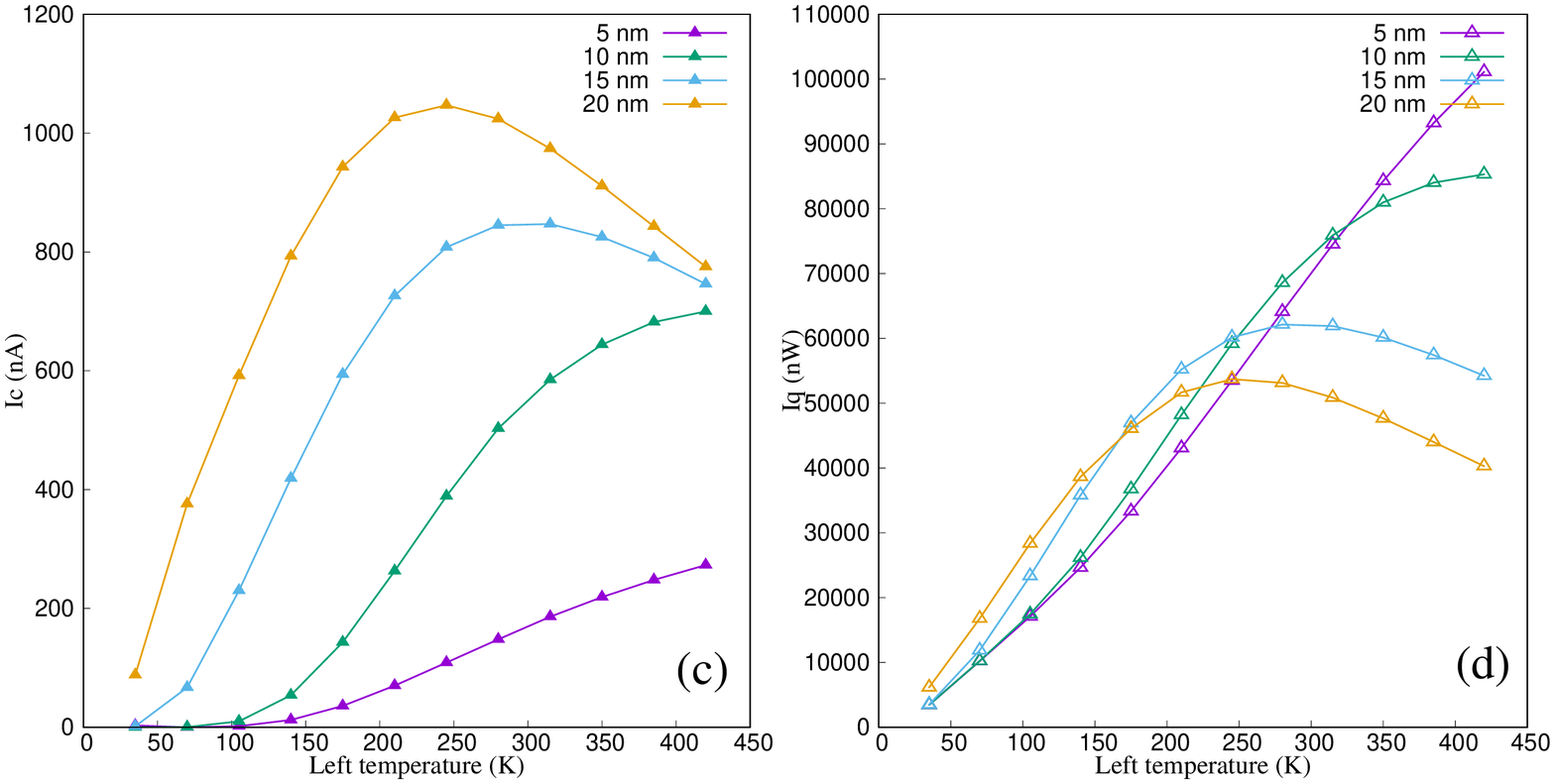}
 \caption{Shell thickness effect on electrical  and heat currents as function of left chemical potential (a) and (b) and temperature (c) and (d). (a) and (c) are representative of electrical current and (b) and (d) are for heat current.   }
 \label{variable_thickness}
\end{figure}

In Figure~\ref{variable_thickness}  we present values of the charge and heat currents for different thickness of the equilateral triangular shell while increasing the left chemical potential and temperature. The higher chemical potential leads more states to participate in transmission and the electric current increase, and this is clearly observable in in Figure~\ref{variable_thickness}(a) regardless of thickness of sample.
But this increase is different for each thickness, 
and from a step like behavior for the 5\,nm shell it evolves to an almost linear function. An equation to describe analytically this I-V characteristics is however beyond the intention of our present study. For thinner shells with respect to the diameter of nanowires, corner localization cause more effect on electrons distribution. And for thin shells the energy gap between the corner and side states are higher. Due to this reason step like behavior was expected from thinner shell and by increasing shell thickness this step like behavior vanish gradually. We can expect these step like behavior for other geometries if we decrease the thickness of shell below 10\,nm.

%
%
%
Fig.~\ref{variable_thickness}(b) shows the variation of the heat current with increasing the left chemical potential, with one or two maxima for each thickness. The energy interval between maxima is related to the energy gap between corner and side states.
Fig.~\ref{variable_thickness}(c) shows that the thermoelectric current  increases with increasing the shell thickness, with an almost linear variation with the temperature bias for the smallest thicknesses, and a with a maximum point for the largest thickness.
Finally, in Fig.~\ref{variable_thickness}(d) we can also see a linear behavior of the heat current for 5\,nm thickness, and a non-linear characteristic with a maximum current developing for the other cases. Interestingly, the relative magnitude of the these heat currents reverses between low and high temperatures. 

\section{Conclusions}

In this paper we reported computed features of charge and heat transport in core-shell nanowires when the electronic transport occurs within the shell. These features were studied for different cross sectional geometries of the shell. The charge and heat conductivity, as functions with respect to the chemical potential and temperature biases, depend significantly on the geometry. Shells with triangular cross section lead to larger charge and heat currents in the presence of a temperature gradient, but not in the case of a chemical potential bias. Also, the charge and heat currents were found to change substantially
with decreasing shell thickness, due to the increasing of the gap between corner and side states.

\section*{Acknowledgment}

This work was supported by funding from the Icelandic Research Fund Grant Nos. 195943-051 and 229078-051.

%
%

\bibliography{Bibliography}

\end{document}